\documentclass[aps,prl,preprint,superscriptaddress,preprintnumbers,nofootinbib,tightenlines,floatfix]{revtex4-1}

\usepackage{graphicx}% Include figure files
\usepackage{dcolumn}% Align table columns on decimal point
\usepackage{bm}% bold math
\usepackage[figuresleft]{rotating}

\newcommand{\dztorhoenu}{D^0 \to \rho^- e^+ \nu_e}
\newcommand{\dptorhoenu}{D^+ \to \rho^0 e^+ \nu_e}
\newcommand{\dtorhoenu}{D \to \rho e^+ \nu_e}

\newcommand{\omegaenu}{D^+ \to \omega e^+ \nu_e}

\newcommand{\dptokstarenu}  {D^+ \to \bar{K}^{*0} e^+ \nu}
\newcommand{\kenu}      {D^0 \to K^- e^+ \nu}

\newcommand{\vcd}{V_{cd}}
\newcommand{\vcs}{V_{cs}}

\newcommand{\dele}{\Delta E}

\newcommand{\eff}{\epsilon}

\newcommand{\psipp}{\psi(3770)}

\newcommand{\pip}{\pi^+}
\newcommand{\pim}{\pi^-}
\newcommand{\piz}{\pi^0}
\newcommand{\pipi}{\pip\pim}

\newcommand{\ra}{\rightarrow}

\newcommand{\ddb}{D\bar{D}}
\newcommand{\ipb}{{\rm pb}^{-1}}

\newcommand{\Dz}{D^{0}}

\newcommand{\Dp}{D^+}
\newcommand{\Dm}{D^-}
\newcommand{\enu}{e^+ \nu_e}

\newcommand{\mbc}{M_{\text{bc}}}
\newcommand{\ks}{K_S^0}

\newcommand{\hp}{H_{+}{(q^2)}}
\newcommand{\hn}{H_{-}{(q^2)}}
\newcommand{\hz}{H_{0}{(q^2)}}

\newcommand{\bmath}{\begin{displaymath}}
\newcommand{\emath}{\end{displaymath}}

\newcommand{\beq}{\begin{equation}}
\newcommand{\eeq}{\end{equation}}
\newcommand{\bfg}{\begin{figure}}
\newcommand{\efg}{\end{figure}}
\newcommand{\bitm}{\begin{itemize}}
\newcommand{\eitm}{\end{itemize}}
\newcommand{\bnum}{\begin{enumerate}}
\newcommand{\enum}{\end{enumerate}}
\newcommand{\btbl}{\begin{table}}
\newcommand{\etbl}{\end{table}}
\newcommand{\btbu}{\begin{tabular}}
\newcommand{\etbu}{\end{tabular}}
\newcommand{\cleoc}{\mbox{CLEO-c}}

\newcommand{\bfdzrhom}{1.77 \pm 0.12 \pm 0.10}
\newcommand{\bfdprhoz}{2.17 \pm 0.12 ^{+0.12}_{-0.22}}
\newcommand{\bfdpomega}{1.82 \pm 0.18 \pm 0.07}

\newcommand{\bfdpomegapmil}{(\bfdpomegapmil) \times 10^{-3}}

\newcommand{\isospinratio}{1.03 \pm 0.09 ^{+0.08}_{-0.02}}

\newcommand{\rv}{1.48 \pm 0.15 \pm 0.05}
\newcommand{\rtwo}{0.83 \pm 0.11 \pm 0.04}

\newcommand{\aonez}{0.56\pm0.01^{+0.02}_{-0.03}}
\newcommand{\atwoz}{0.47\pm0.06\pm0.04}
\newcommand{\vz}{0.84\pm0.09^{+0.05}_{-0.06}}

\begin{document}

\title{\boldmath First Measurement of the Form Factors in the Decays $D^0 \rightarrow \rho^- e^+ \nu_e $ and $D^+ \ra\rho^0\enu$}

\preprint{CLNS 11/2075}  % the CLNS number
\preprint{CLEO 11-3 \hfill}    % the CLEO number

\author{S.~Dobbs}
\author{Z.~Metreveli}
\author{K.~K.~Seth}
\author{A.~Tomaradze}
\author{T.~Xiao}
\affiliation{Northwestern University, Evanston, Illinois 60208, USA}
\author{L.~Martin}
\author{A.~Powell}
\author{G.~Wilkinson}
\affiliation{University of Oxford, Oxford OX1 3RH, UK}
\author{H.~Mendez}
\affiliation{University of Puerto Rico, Mayaguez, Puerto Rico 00681}
\author{J.~Y.~Ge}
\author{G.~S.~Huang}
\altaffiliation[Now at: ]{University of Science and Technology of China, Hefei 230026, People's Republic of China}
\author{D.~H.~Miller}
\author{V.~Pavlunin}
\altaffiliation[Now at: ]{University of California, Santa Barbara, CA 93106, USA}
\author{I.~P.~J.~Shipsey}
\author{B.~Xin}
\affiliation{Purdue University, West Lafayette, Indiana 47907, USA}
\author{G.~S.~Adams}
\author{D.~Hu}
\author{B.~Moziak}
\author{J.~Napolitano}
\affiliation{Rensselaer Polytechnic Institute, Troy, New York 12180, USA}
\author{K.~M.~Ecklund}
\affiliation{Rice University, Houston, Texas 77005, USA}
\author{J.~Insler}
\author{H.~Muramatsu}
\author{C.~S.~Park}
\author{L.~J.~Pearson}
\author{E.~H.~Thorndike}
\affiliation{University of Rochester, Rochester, New York 14627, USA}
\author{S.~Ricciardi}
\affiliation{STFC Rutherford Appleton Laboratory, Chilton, Didcot, Oxfordshire, OX11 0QX, UK}
\author{C.~Thomas}
\affiliation{University of Oxford, Oxford OX1 3RH, UK}
\affiliation{STFC Rutherford Appleton Laboratory, Chilton, Didcot, Oxfordshire, OX11 0QX, UK}
\author{M.~Artuso}
\author{S.~Blusk}
\author{R.~Mountain}
\author{T.~Skwarnicki}
\author{S.~Stone}
\author{L.~M.~Zhang}
\affiliation{Syracuse University, Syracuse, New York 13244, USA}
\author{G.~Bonvicini}
\author{D.~Cinabro}
\author{A.~Lincoln}
\author{M.~J.~Smith}
\author{P.~Zhou}
\author{J.~Zhu}
\affiliation{Wayne State University, Detroit, Michigan 48202, USA}
\author{P.~Naik}
\author{J.~Rademacker}
\affiliation{University of Bristol, Bristol BS8 1TL, UK}
\author{D.~M.~Asner}
\altaffiliation[Now at: ]{Pacific Northwest National Laboratory, Richland, WA 99352}
\author{K.~W.~Edwards}
\author{K.~Randrianarivony}
\author{G.~Tatishvili}
\altaffiliation[Now at: ]{Pacific Northwest National Laboratory, Richland, WA 99352}
\affiliation{Carleton University, Ottawa, Ontario, Canada K1S 5B6}
\author{R.~A.~Briere}
\author{H.~Vogel}
\affiliation{Carnegie Mellon University, Pittsburgh, Pennsylvania 15213, USA}
\author{P.~U.~E.~Onyisi}
\author{J.~L.~Rosner}
\affiliation{University of Chicago, Chicago, Illinois 60637, USA}
\author{J.~P.~Alexander}
\author{D.~G.~Cassel}
\author{S.~Das}
\author{R.~Ehrlich}
\author{L.~Gibbons}
\author{S.~W.~Gray}
\author{D.~L.~Hartill}
\author{B.~K.~Heltsley}
\author{D.~L.~Kreinick}
\author{V.~E.~Kuznetsov}
\author{J.~R.~Patterson}
\author{D.~Peterson}
\author{D.~Riley}
\author{A.~Ryd}
\author{A.~J.~Sadoff}
\author{X.~Shi}
\author{W.~M.~Sun}
\affiliation{Cornell University, Ithaca, New York 14853, USA}
\author{J.~Yelton}
\affiliation{University of Florida, Gainesville, Florida 32611, USA}
\author{P.~Rubin}
\affiliation{George Mason University, Fairfax, Virginia 22030, USA}
\author{N.~Lowrey}
\author{S.~Mehrabyan}
\author{M.~Selen}
\author{J.~Wiss}
\affiliation{University of Illinois, Urbana-Champaign, Illinois 61801, USA}
\author{J.~Libby}
\affiliation{Indian Institute of Technology Madras, Chennai, Tamil Nadu 600036, India}
\author{M.~Kornicer}
\author{R.~E.~Mitchell}
\author{C.~M.~Tarbert}
\affiliation{Indiana University, Bloomington, Indiana 47405, USA }
\author{D.~Besson}
\affiliation{University of Kansas, Lawrence, Kansas 66045, USA}
\author{T.~K.~Pedlar}
\affiliation{Luther College, Decorah, Iowa 52101, USA}
\author{D.~Cronin-Hennessy}
\author{J.~Hietala}
\affiliation{University of Minnesota, Minneapolis, Minnesota 55455, USA}
\collaboration{CLEO Collaboration}
\noaffiliation

\date{December 13, 2011}

\begin{abstract}

Using the entire CLEO-c $\psipp\to\ddb$ event sample, corresponding to an integrated luminosity of 818 $\ipb$
and approximately 5.4 $\times$ 10$^6$ $\ddb$ events, 
we measure the form factors for the decays
$\dztorhoenu$ and $\dptorhoenu$  for the first time 
and the branching fractions with improved precision.
A four-dimensional unbinned maximum likelihood fit determines the form factor ratios to be:
$V(0)/A_1(0) = \rv$ and $A_2(0)/A_1(0)= \rtwo$.
Assuming CKM unitarity, the known $D$ meson lifetimes and our measured branching fractions we obtain the form factor normalizations $A_1(0)$, $A_2(0)$, and $V(0)$. We also present a measurement of the branching fraction for $\omegaenu$ with improved precision.
\end{abstract}

\maketitle

The transition rate of charm semileptonic decays depends on the weak quark mixing
Cabibbo-Kobayashi-Maskawa (CKM) matrix elements
$|\vcs|$ and $|\vcd|$~\cite{ckm}, and strong interaction effects binding
quarks into hadrons parameterized by form factors.

In the decays $\dtorhoenu$,
in the limit of negligible lepton mass,
the hadronic current is described by three dominant
form factors: two axial and one vector, $A_1$, $A_2$, and $V$, respectively,
which are functions of $q^2$, the invariant mass of the lepton-neutrino system.
They are not amenable to unquenched LQCD calculations due to the large
total decay width of the $\rho$ meson, but model predictions
exist~\cite{isgw2,fajfer}. No experimental information on these form
factors exists.

The helicity amplitudes for the rare decays $B \to V \ell^+\ell^-$ are related
at leading order in $\Lambda_{\rm QCD}/m_b$ to pseudoscalar-to-vector semileptonic
transitions~\cite{grinst2}.
Exploiting one of the proposed double-ratio techniques~\cite{doubleratio},
$D\to\rho\enu$ form factors,
when combined with those of $D\to K^{*} \enu$ and $B \to V \ell^+\ell^-$,
can be used to extract $|V_{ub}|$ from $B \to \rho \enu$.

The differential decay rate of $\dtorhoenu$ can be expressed in terms of three helicity amplitudes
($\hp$, $\hn$, and $\hz$)~\cite{ptovtheory}:
\begin{widetext}
\begin{eqnarray}
\frac{d \Gamma}{dq^2\, d\!\cos{\theta_{\pi}}\, d\!\cos{\theta_e}\, d\chi\, dm_{\pi\pi}} = \hspace{10cm}
\nonumber \\
\frac{3 }{8(4 \pi)^4}G_F^2 |V_{cd}|^2 \frac{p_{\rho}q^2}{M_D^2}\mathcal{B}(\rho \rightarrow \pi \pi) |{\cal BW}(m_{\pi\pi})|^2
   \Big[ (1+\cos{\theta_e})^2 \sin^2{\theta_\pi} |H_+(q^2,m_{\pi\pi})|^2  \nonumber \\
 \hspace{0.5cm} + (1-\cos{\theta_e})^2 \sin^2{\theta_\pi} | H_-(q^2,m_{\pi\pi}) |^2
    + 4 \sin^2 \theta_e \cos^2 \theta_\pi | H_0(q^2,m_{\pi\pi}) |^2 \nonumber \\
 \hspace{0.5cm} + 4 \sin \theta_e (1+\cos \theta_e) \sin \theta_\pi \cos \theta_\pi \cos \chi H_+(q^2,m_{\pi\pi}) H_0(q^2,m_{\pi\pi}) \nonumber \\
 \hspace{0.5cm} - 4 \sin \theta_e (1-\cos \theta_e) \sin \theta_\pi \cos \theta_\pi \cos \chi H_-(q^2,m_{\pi\pi}) H_0(q^2,m_{\pi\pi}) \nonumber \\
 \hspace{0.5cm} - 2 \sin^2 \theta_e \sin^2 \theta_\pi \cos 2\chi H_+(q^2,m_{\pi\pi}) H_-(q^2,m_{\pi\pi}) \Big], %\nonumber
\label{eq:decayrate}
\end{eqnarray}
\end{widetext}
where $G_F$ is the Fermi constant,
$p_{\rho}$ is the momentum of the $\rho$ in the $D$ rest frame,
$\mathcal{B}(\rho \rightarrow \pi \pi)$ is a branching fraction,
$\theta_\pi$ is
the angle between the $\pi$ and the $D$ direction in the $\rho$ rest
frame,
$\theta_e$ is
the angle between the $e^+$ and the $D$ direction in the $\enu$ rest frame,
$\chi$ is the acoplanarity angle between the $\pipi$ and $\enu$ decay planes,
$m_{\pi\pi}$ is the invariant mass of the two pions,
and
${\cal BW}(m_{\pi \pi})$ is the Breit-Wigner function that describes the $\rho$ line shape.
Following Ref.~\cite{focusKstarspec},
we use the relativistic form 
\beq
{\cal BW}(m_{\pi\pi}) = \frac{ \sqrt{m_0\Gamma_0}(p/p_0) }{m_0^2 - m_{\pi\pi}^2 - i m_0 \Gamma(m_{\pi\pi}) } \frac {B(p)} {B(p_0)},
\label{eq:bw}
\eeq
where $m_0$ and $\Gamma_0$ are the mass and width of the $\rho$ 
meson~\cite{pdg2010},
$p$ is the momentum of the pion in the $\pi\pi$ rest frame,
$p_0$ is equal to $p$ when $m_{\pi\pi} = m_0$,
and
$B(p)$ is a Blatt-Weisskopf form factor given by
$B(p) = {1}/{{(1+R^2 p^2)}^{1/2}}$,
with $R = 3$~GeV$^{-1}$, and 
$\Gamma(m_{\pi \pi}) = ({p}/{p_0})^3  ({m_0}/{m_{\pi \pi}}) \Gamma_0 [{B(p)}/{B(p_0)}]^2$.
The interference term between a possible $s$-wave $\pi\pi$ component and the $\rho$ amplitude has not been included in Eq.~(\ref{eq:decayrate}). Its absence is treated as a source of systematic uncertainty on the measurement.

The helicity amplitudes are related to the form factors 
\begin{eqnarray}
H_{\pm}(q^2) & = &   M A_1(q^2) \mp 2 \frac{M_D p_{\rho} } { M } V(q^2), \\
H_{0}(q^2)   &  = &  \frac{1}{2 m_{\pi \pi} \sqrt{q^2}} \Big[ (M_D^2 - m_{\pi \pi}^2 - q^2) M A_1(q^2)  \nonumber \\
 & &  - 4 \frac{M_D^2 p^2_{\rho}}{M} A_2(q^2)\Big],
\end{eqnarray}
where $M_D$ is the mass of the $D$ meson and $M=M_D+m_{\pi\pi}$.
Since $A_1(q^2)$ is common to all three helicity
amplitudes, %with limited experimental statistics,
it is natural to define two form factor ratios as
\begin{eqnarray}
r_V = \frac{V(0)}{A_1(0)}\; {\rm and} \; r_2 = \frac{A_2(0)}{A_1(0)}. %\nonumber
\label{eq:rvr2}
\end{eqnarray}
We assume a simple pole form~\cite{simplepole}
 for $A_1(q^2)$, $A_2(q^2)$, and $V(q^2)$,
where the pole mass is $M_{D^*(1^-)}$= 2.01~GeV$/c^2$ and $M_{D^*(1^+)}$= 2.42~GeV$/c^2$~\cite{pdg2010}
 for the vector and axial form factors, respectively.
We have also explored a double-pole parametrization~\cite{fajfer}.

We report herein the first measurement of the form factor ratios and absolute form factor normalization
in $\dtorhoenu$,
and improved branching fraction measurements for these decays and $\omegaenu$.
(Throughout this Letter charge-conjugate modes are implied.)
These decays were studied previously using a smaller \cleoc\ data sample~\cite{56pbsemil}.
The data sample used here consists of an integrated luminosity of
$818~\text{pb}^{-1}$ at the $\psipp$ resonance, and includes about
$3.0\times 10^6$ $D^0\bar{D}^0$ and
$2.4\times 10^6$ $\Dp\Dm$ events.
The CLEO-c detector is
described in detail elsewhere~\cite{cleo_detector}.

The analysis technique was employed in previous \cleoc\ studies~\cite{tagpenu, 56pbsemil}.
The presence of two $D$ mesons in a $\ddb$ event allows
a tag sample to be defined in which a $\bar{D}$
is reconstructed in a hadronic decay mode.
A sub-sample is then formed in which a positron and a set of hadrons, as a signature
of a semileptonic decay, are required in addition to the tag.
The semileptonic decay branching fraction ${\cal{B}}_{{\rm SL}}$ is given by
\begin{equation}
{\cal{B}}_{{\rm SL}} =  \frac {N_{ \rm tag, SL }} {N_{ \rm tag}}   \frac {\epsilon_{ \rm tag}} {\epsilon_{\rm tag, SL}} =
\frac {N_{ \rm tag, SL} / \epsilon} {N_{ \rm tag}},
\label{eq:master}
\end{equation}
where
$N_{\rm tag}$ and $\epsilon_{\rm tag}$ are the yield and reconstruction efficiency, respectively, for the hadronic tag,
$N_{{\rm tag, SL}}$ and $\epsilon_{{\rm tag, SL}}$ are those for the combined semileptonic decay and hadronic tag,
and
$\epsilon = \epsilon_{ \rm tag, SL}/\epsilon_{ \rm tag }$ is the effective signal efficiency.

Candidate events are selected by reconstructing a $\bar{D}^0$ or
$D^-$ tag in the following hadronic final states: $K^+ \pi^-$,
$K^+ \pi^- \pi^0$, 
and $K^+ \pi^- \pi^- \pi^+$
for neutral tags, and $K^0_S \pi^-$, $K^+
\pi^- \pi^-$, $K^0_S \pi^- \pi^0 $, $K^+ \pi^- \pi^- \pi^0$,
$K^0_S \pi^- \pi^- \pi^+$, and $K^- K^+ \pi^- $ for charged tags.
Tagged events are selected based on two variables: $\dele \equiv
E_D - E_{\text{beam}}$, the difference between the energy of the
$\Dm$ tag candidate $E_D$ and the beam energy
$E_{\text{beam}}$, and the beam-constrained mass $\mbc \equiv
(E^2_{\text{beam}}/c^4 - |{\bf p}_D|^2/c^2)^{1/2}$, where
${\bf p}_D$ is the measured momentum of the $\Dm$ candidate.
Selection criteria for tracks, $\piz$, and $\ks$ candidates used in
the reconstruction of tags are described in Ref.~\cite{dhad281}.
If multiple candidates are present in the same tag mode, one
candidate per tag charge with the smallest $|\Delta E|$ is chosen.
The yield of each tag mode is obtained from fits to the $\mbc$ distributions~\cite{dhad281}.
The
data sample comprises 661232$\pm$879 and 481927$\pm$810 reconstructed neutral and charged
tags, respectively.

After a tag is identified, we search for an $e^+$ and
a $\rho^-$ ($\pim\piz$ mode), $\rho^0$ ($\pipi$ mode), or $\omega$ ($\pipi\piz$ mode)
recoiling against the tag following Ref.~\cite{dhad281}.
A $\rho\ra\pi\pi$ candidate satisfies $|m_{\pi\pi}-m_0|<$ 150 MeV$/c^2$.
The combined tag and semileptonic candidates must account for all tracks in the event.
Semileptonic decays are identified with
%using
%the variable
$U \equiv E_{\text{miss}} - c|{\bf p}_{\text{miss}}|$, where
$E_{\text{miss}}$ and ${\bf p}_{\text{miss}}$ are the missing energy and
momentum of the $D^+$ meson. If the decay products have been correctly identified,
$U$ is expected to be zero, since only a neutrino is undetected.
The resolution in $U$ is improved
by constraining the magnitude and direction of the $D^+$ momentum to be
$p_{D^+} =(E^2_{\rm beam}/c^2-c^2m^2_{D})^{1/2}$, and
$\widehat{\bf p}_{D^+} = - \widehat{\bf p}_{D^-}$~\cite{56pbsemil}, respectively.
Due to the finite resolution of the detector,
the distribution in $U$ is approximately Gaussian,
with resolution $\sim$17~MeV for $\dztorhoenu$ and $\omegaenu$
and $\sim$8~MeV for $\dptorhoenu$.
To
remove multiple candidates in each semileptonic mode one
combination is chosen per tag mode per tag charge, based on the
proximity of the invariant masses of the $\rho^0$, $\rho^+$,
or $\omega$ candidates to their expected masses.

\begin{figure}[tb]
\includegraphics*[width=5.in]{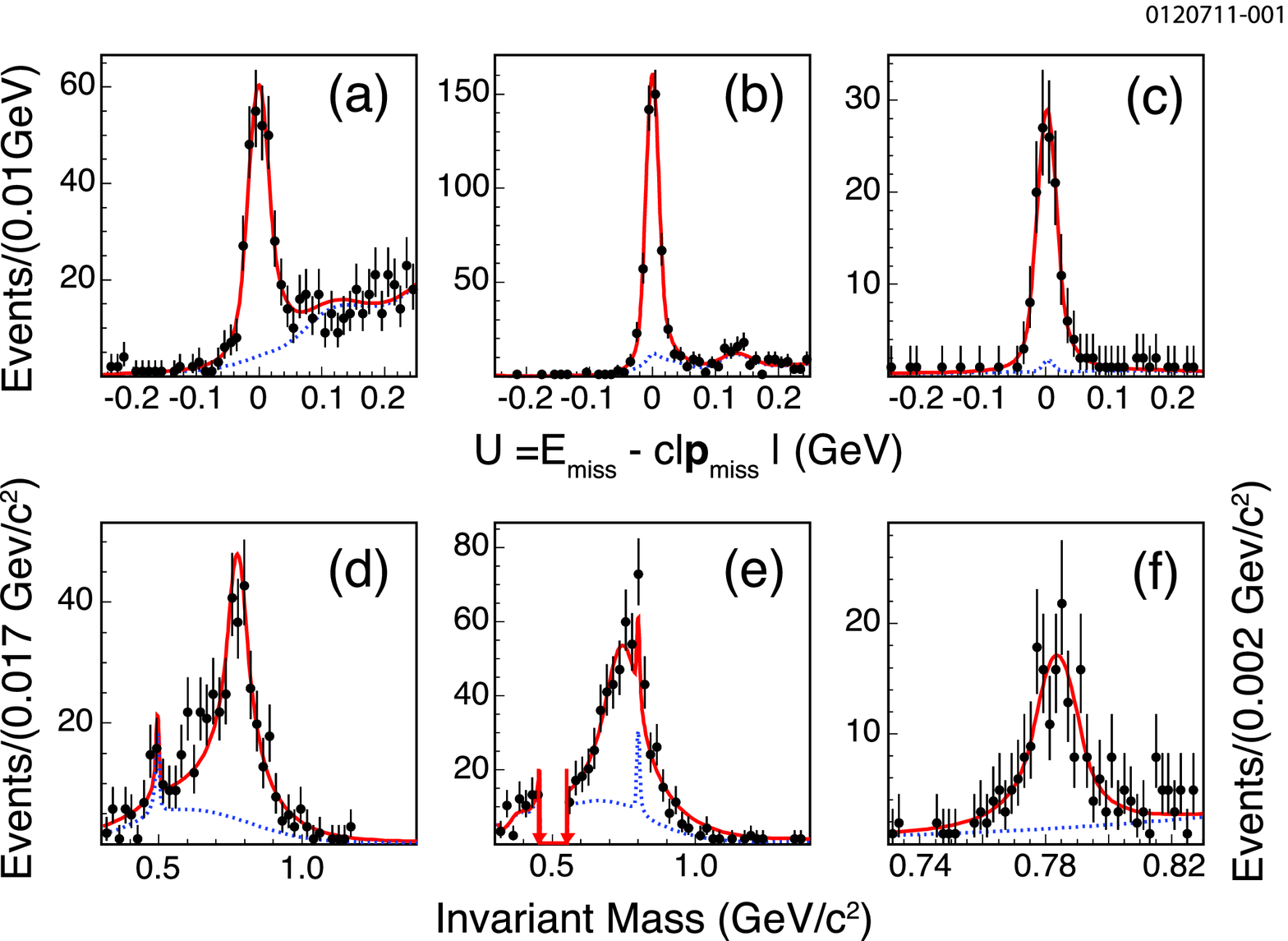}
\caption{Fits to the $U$ and hadron invariant mass 
distributions in data (filled circles
with error bars) for
(a) and (d) $\dztorhoenu$, $\rho^-\ra\pim\piz$;
(b) and (e) $\dptorhoenu$, $\rho^0\ra\pipi$; and
(c) and (f) $\omegaenu$, $\omega\ra\pipi\piz$.
The solid line
represents the fit of the sum of the signal function and
background function to the data. The dashed line indicates the
background contribution. 
The arrows indicate a $\pm$48 MeV region  around the $\ks$ mass, which has been removed for display.
} \label{fig:U}
\end{figure}

The $U$ and invariant mass distributions for $\dztorhoenu$, $\dptorhoenu$,
and $\omegaenu$ with all tag modes
combined are shown in Fig.~\ref{fig:U}.
The
yield for each of the three modes is determined
from a binned likelihood
fit to the $U$ distribution where the signal is
described by a modified Crystal Ball function with two power-law
tails~\cite{cb_2tail} which account for initial- and final-state
radiation (FSR) and mismeasured tracks. The signal parameters
are fixed with a GEANT-based Monte Carlo (MC) simulation~\cite{geant} in fits to the data.
The background functions are determined
by MC simulation
that incorporates all available data on $D$ meson
decays, which we refer to as \lq\lq generic MC".
For $\dztorhoenu$, the backgrounds arise mostly from $D^0 \rightarrow K^{*-} e^+ \nu_e$,
peaking at positive $U$ and modeled with a Gaussian, and events with misidentified tags,
which are accounted for in the fit by
a fourth order polynomial.
The backgrounds to $D^+ \rightarrow \rho^0 e^+ \nu_e$ has its largest contribution from
$\dptokstarenu$, $\bar{K}^{*0}\to K^-\pi^+$,
with the peak at higher $U$ due to 
charged kaons misidentified as charged pions,
and the peak at lower $U$ from either decay-in-flight kaons or
interactions with detector material. 
We categorize the background components
according to their shape in $U$ and parameterize the overall background shape using combinations of polynomials and Gaussian functions. 
The background shape parameters are fixed in fits to the data, while the background normalizations are allowed
to float.
The signal shapes for the invariant mass distributions of the hadronic system
are modeled with a Breit-Wigner function, and the background shapes are modeled with generic MC.
The peaking background for $\dptorhoenu$ arises from $\omegaenu$, $\omega\to\pipi$.
Due to the tag, backgrounds
from the non-$D \bar{D}$ processes $e^+ e^- \rightarrow q
\bar{q}$, where $q$ is a $u$, $d$, or $s$ quark, $e^+ e^-
\rightarrow \tau^+ \tau^-$, and $e^+ e^- \rightarrow \psi(2S)
\gamma$, are negligible~\cite{tagpenu}.
The signal yields $N_{\rm tag, SL}$ are given in Table~\ref{tab:br}.
%The fits describe the data well.

The second row of Fig.~\ref{fig:U} shows the $m_{\pim\piz}$,
$m_{\pipi}$, and $m_{\pipi\piz}$ 
distributions with $|U|<$ 60~MeV for the three signal modes, respectively. 
The peaking background at $m_{\pim\piz}\sim$ 0.49 GeV$/c^2$ arises
from $\kenu$ with $K^-\ra\pim\piz$.
The small background peak at  $m_{\pipi}\sim$ 0.78 GeV$/c^2$ is due to
$\omegaenu$ with $\omega\ra\pipi$~\cite{rhoomegafn}. 

\begingroup
\squeezetable
\begin{table*}[tb]
\caption{Signal efficiencies, yields, and
branching fractions (${\cal B}_{\rm SL}$) for $\dztorhoenu$, $\dptorhoenu$, and $\omegaenu$,
from this work, our previous (prev) measurements~\cite{56pbsemil}, and two model 
predictions: ISGW2~\cite{isgw2} and FK~\cite{fajfer}.
All ${\cal B}_{\rm SL}$ are in units of $10^{-3}$. 
The uncertainties for $\eff$ and $N_{\text{tag, SL}}$
are statistical, while the uncertainties for branching fractions are
statistical and systematic in that order. 
The efficiencies include the $\rho$ and $\omega$ decay branching fractions  from the PDG~\cite{pdg2010}.
}
\begin{center}
\begin{tabular}{lccccccc}\hline\hline
Decay Mode   &$\eff$ (\%) &  $N_{\text{tag, SL}} $ & ${\cal B}_{\rm SL} $ &
$~{\cal B}_{\rm SL}$(prev)& $~{\cal B}_{\rm SL} $(ISGW2) & $~{ \cal B}_{\rm SL}$(FK)
\rule[-1mm]{0mm}{4.3mm}   \\
\hline
$\Dz\ra\rho^-\enu$~~&26.03 $\pm$ 0.02 &  ~304.6 $\pm$ 20.9~ & ~$\bfdzrhom$~ & ~1.94 $\pm$ 0.39 $\pm$ 0.13~& ~1.0 ~& ~2.0~ \rule[-1mm]{0mm}{4.3mm}\\
$\Dp\ra\rho^0\enu$~~&  42.84 $\pm$ 0.03   &     447.4 $\pm$ 24.5    &    $\bfdprhoz$    &    2.1 $\pm$ 0.4 $\pm$ 0.1      &    1.3    & 2.5       \\
$\Dp\ra\omega\enu$~~&  14.67 $\pm$ 0.03   &     128.5 $\pm$ 12.6    & $\bfdpomega$    & $1.6^{+0.7}_{-0.6} \pm$ 0.1& 1.3    & 2.5       \\
\hline \hline
\end{tabular}
\end{center}
\label{tab:br}
\end{table*}
\endgroup

The absolute branching fractions in Table~\ref{tab:br} are obtained using
Eq.~(\ref{eq:master}). The signal efficiencies $\epsilon$ are determined by MC simulation, and have been weighted
by the tag yields in the data.

The systematic uncertainties for the branching fractions of $\dztorhoenu$ and $\dptorhoenu$ are dominated by uncertainties
in the line shape of the $\rho$ (5.0\%), 
and the non-resonant background ($-$1.5\% for $\dztorhoenu$ and
$-$8.4\% for $\dptorhoenu$).
The uncertainty due to the line shape of the $\rho$ is estimated by 
(1) requiring $|U|<$ 60~MeV and fitting the $m_{\pi\pi}$ distribution,
(2) varying the selection criterion $|m_{\pi\pi}-m_0|<$ 150~MeV.
The uncertainty due to the non-resonant background is obtained by
performing a form factor fit, 
with an additional interfering non-resonant $D\to\pi\pi\enu$ ($s$-wave) component modeled following Ref.~\cite{swavefocus},
then integrating over the kinematic variables to recalculate the branching fractions.
The unknown form factors in $\omegaenu$
are the dominant uncertainty in its branching fraction (3.0\%).
The remaining systematic uncertainties include
the track and $\piz$ finding efficiencies, positron and charged hadron identification,
the number of tags, the no-additional-track requirement,
the shape of the signal and background functions,
and the MC FSR and form factor modeling.
These estimates are added in quadrature
to obtain the total systematic uncertainties on the branching fractions:
$^{+5.7}_{-5.9}$\%, $^{~+5.5}_{-10.0}$\%, 4.1\%, for $\dztorhoenu$, $\dptorhoenu$, and $\omegaenu$, respectively.

A form factor analysis is performed for $\dtorhoenu$.
We calculate the energy and momentum of the neutrino using
$E_\nu  =  E_{\rm miss}$ and $|{\bf p}_{\nu}|  =  E_{\rm miss}$,
because $E_{\rm miss}$ is better measured than $|{\bf p}_{\rm miss}|$.
Without ambiguity, the four kinematic variables 
($q^2$, $\cos\theta_\pi$, $\cos\theta_e$ $\chi$) are measured
with resolutions of (0.021~GeV$^2/c^4$, 0.020, 0.048, 0.024) for $\dztorhoenu$,
and (0.013~GeV$^2/c^4$, 0.013, 0.037, 0.019) for $\dptorhoenu$.

A four-dimensional maximum likelihood fit in a manner similar to Ref.~\cite{witherell} is performed in the space
of $q^2$, $\cos\theta_\pi$, $\cos\theta_e$, and $\chi$.
The technique makes possible a multidimensional fit to variables modified by experimental acceptance and resolution taking into account correlations among the variables.
The signal probability density function for the likelihood function is estimated
at each data point using signal MC events by sampling the MC distribution
at the reconstructed level in a search volume around the data point,
then weighting by the ratio of the decay distribution for the trial values of $r_V$ and $r_2$ to that of the generated distribution.
The search volumes are one tenth
 the full kinematic range of each of the four dimensions.
Large MC samples are generated to ensure that each search volume has sufficient statistics.
The background probability density function is modeled using events from the generic MC.
Due to the low statistics of the background in the generic MC, we reduce
the four dimensional space to lower dimensional subspaces.
Due to the correlation between $q^2$ and $\cos\theta_e$, the two
subspaces are chosen to be ($q^2$, $\cos\theta_e$) and ($\cos\theta_\pi$, $\chi$).
The background normalization is fixed in the fits to the values measured in the determination of the branching fractions.

\begin{figure}[tb]
\includegraphics*[width=5.in]{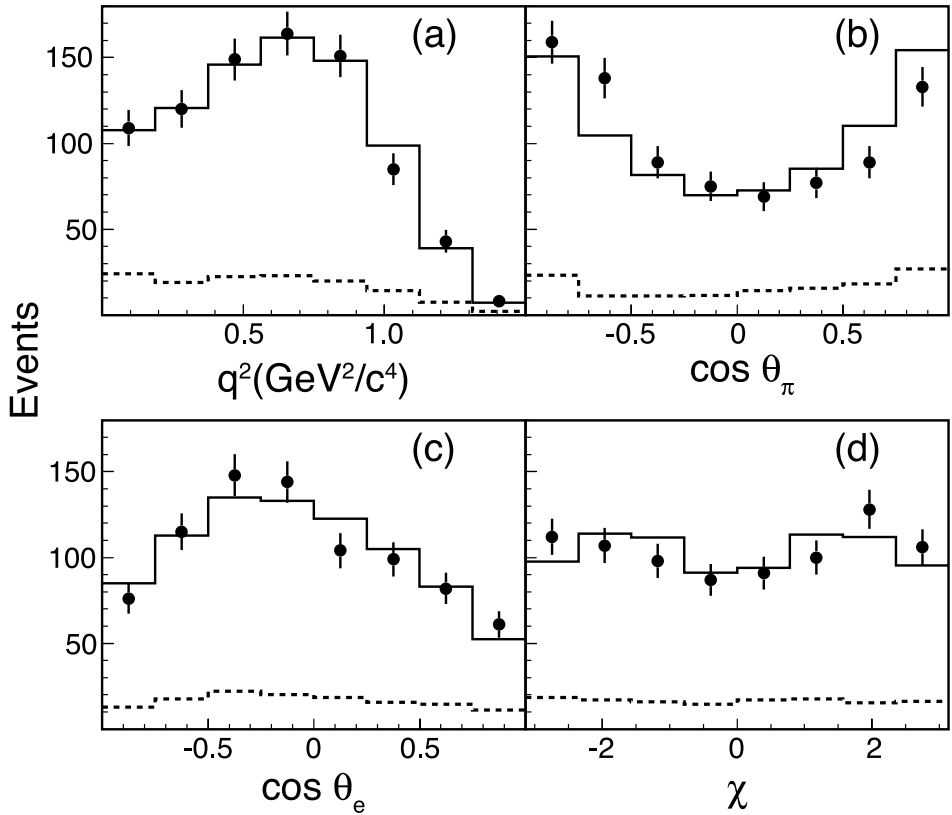}
\caption{Projections of the combined $\rho^-$ and $\rho^0$ data
  (points with statistical error bars) 
and the fit (solid histogram) onto
$q^2$, $\cos\theta_e$, $\cos\theta_\pi$, and $\chi$.
The dashed lines show the sum of the background distributions.
} \label{fig:proj}
\end{figure}

Using the above method,
a simultaneous fit is made to the isospin-conjugate modes $\dztorhoenu$ and $\dptorhoenu$.
We find $r_V = 1.48 \pm 0.15$ and $r_2 = 0.83 \pm 0.11$,
with a correlation coefficient $\rho_{V2} = -$0.18.
The confidence level of the fit is determined to be 5.0\% by comparing
the negative log-likelihood from the data to the distribution from toy
MC fits.
Fig.~\ref{fig:proj} shows the $q^2$, $\cos\theta_e$, $\cos\theta_\pi$,
and $\chi$ projections for the combined $\rho^-$ and $\rho^0$ data and
the fit.
We also make fits to the two modes separately. The results are consistent.
We note that the difference between the data and the fit projection
for $\cos\theta_\pi$ might be due to $s$-wave interference.

We have considered the following sources of systematic uncertainty in the form factor measurement.
Our estimate of their magnitude are given in parentheses for $r_V$ and $r_2$, respectively.
The uncertainty associated with background modeling (0.01, 0.02) is estimated by
changing the normalization of the three largest background components by a factor of two in each semileptonic mode.
The uncertainty due to imperfect knowledge of the $\rho$ line shape (0.01, 0.02) is estimated by
modifying the $\rho$ line shape by
increasing and decreasing the population of signal MC events
below and above the nominal $\rho$ mass~\cite{pdg2010} by 20\%.
The uncertainty due to non-resonant background (0.01, 0.02) is obtained by
repeating the fit
with an additional interfering non-resonant $D\to\pi\pi\enu$ component
($s$-wave) following Ref.~\cite{swavefocus}.
The procedure for extracting the form factor parameters is
tested using the generic MC sample, from which events are drawn randomly to form mock data samples,
each equivalent in size to the data sample.
When backgrounds are absent, the measured form factor ratios are
consistent with the input values.
In the presence of background, a small statistically significant shift is observed. 
Its magnitude is taken as the uncertainty
due to possible bias in the form factor fitter (0.03, 0.02).
The uncertainty associated with the unknown $q^2$ dependence of the form factors (0.03, 0.02) is estimated by
introducing a second pole~\cite{fajfer}.

Adding all sources of systematic uncertainty in quadrature, the final result is
$r_V = \rv$ and $r_2 = \rtwo$.
Using $|V_{cd}| = 0.2252 \pm 0.0007$
obtained using CKM unitarity constraints~\cite{pdg2010}
and the lifetimes
$\tau_{D^0}=(410.1 \pm 1.5)\times 10^{-15} {\rm s}$
and $\tau_{D^+}= (1040 \pm 7)\times 10^{-15}{\rm s}$~\cite{pdg2010},
we combine our form factor ratio and branching fraction results
to obtain
$A_1(0)= \aonez$, $A_2(0)=\atwoz$, and $V(0)=\vz$.

Our branching fraction results are compared to previous measurements~\cite{56pbsemil}, 
with which they are consistent, and theoretical predictions in Table~\ref{tab:br}. 
The results are consistent with isospin invariance:
$\frac{\Gamma(D^0\to\rho^-\enu)}{2\Gamma(D^+\to\rho^0\enu)} =
\isospinratio$.
Isospin symmetry is not expected to be exact due to $\rho^0-\omega$ interference~\cite{rhoomegafn}.
Theoretical predictions
from the ISGW2 model~\cite{isgw2} and a model (FK) which combines
heavy-quark symmetry and properties of the chiral
Lagrangian~\cite{fajfer}, are also listed in Table~\ref{tab:br}.
The branching fractions for ISGW2 are obtained by combining the partial rates in Ref.~\cite{isgw2}
with $|V_{cd}|$ and $\tau_D$ from PDG~\cite{pdg2010}.
Our branching fraction results are more consistent with the FK predictions
than ISGW2.

The FK model predicts
$A_1(0)=0.61$, $A_2(0)=0.31$, and $V(0)=1.05$.
These values are compatible with our form factor measurements.
No other experimental form factor results on these decays exist. %to date.
Our values of $r_V$ and $r_2$ are very similar to
the current PDG average of $\dptokstarenu$ form factor ratios
$r_V=1.62\pm 0.08$ and $r_2 = 0.83\pm 0.05$~\cite{pdg2010}.

In summary, we have made the first measurement of the form factor
ratios and absolute form factor normalization in $\dtorhoenu$,
and improved branching fraction measurements for these decays and $\omegaenu$.
Our branching fractions are consistent with our previous measurements but with improved precision.
The form factor measurement in $\dtorhoenu$ is the first in
a semileptonic Cabibbo-suppressed pseudoscalar-to-vector transition.

\begin{acknowledgments} 
We gratefully acknowledge the effort of the CESR staff 
in providing us with excellent luminosity and running conditions. 
This work was supported by the 
A.P.~Sloan Foundation, 
the National Science Foundation, 
the U.S. Department of Energy, 
the Natural Sciences and Engineering Research Council of Canada, and 
the U.K. Science and Technology Facilities Council. 
\end{acknowledgments}


\begin{thebibliography}{99}
\bibitem{rhoomegafn} $\rho^0-\omega$ interference
is not included in the $\rho^0$ line shape, but is accounted for both by
fitting the $U$ distribution to estimate and subtract the $\omegaenu$
with $\omega\ra\pipi$ contribution and by large variations in the
$\rho^0$ (and $\rho^+$) line shape in the determination of the
branching fractions and form factors.
\end{thebibliography}
\end{document}